\documentclass{article}
\usepackage{spconf,amsmath,graphicx}

\usepackage{multirow}

\usepackage{color}
\usepackage{enumitem}

\def\etal{\emph{et al.}}

\definecolor{mygreen}{RGB}{0,120,0}
\definecolor{myred}{RGB}{120,0,0}
\definecolor{blueViolet}{RGB}{5,52,163}

\usepackage{amsmath}
\usepackage{amsfonts}
\usepackage{amssymb}
\usepackage{amsthm}
\usepackage{textcomp}

\definecolor{redStrong}{rgb}{.85,0,0}
\definecolor{greenStrong}{rgb}{0,.5,0}

\newcommand{\mdC}[1]{  #1} 

\DeclareMathOperator*{\argmin}{arg\,min\,}
\DeclareMathOperator*{\argmax}{arg\,max\,}

\newcommand{\K}{K}

\newcommand{\bx}{\mathbf{x}}
\newcommand{\bz}{\mathbf{z}}
\newcommand{\z}{z}

\renewcommand{\u}{u}

\newcommand{\bomega}{\pmb{\omega}}

\newcommand{\bS}{\mathbf{S}}

\newcommand{\bI}{\mathbf{I}}
\newcommand{\bF}{\mathbf{F}}

\newcommand{\bB}{\mathbf{B}}
\newcommand{\bW}{\mathbf{W}}

\newcommand{\btau}{{\pmb{\tau}}}
\newcommand{\bSigma}{{\pmb{\Sigma}}}

\newcommand{\bGamma}{{\pmb{\Gamma}}}

\usepackage{tikz}
\graphicspath{{images/}{images/realData/}}

\title{A Practical Guide to Multi-image Alignment}
\name{Cecilia Aguerrebere$^{\dagger}$ \qquad Mauricio Delbracio$^{\star}$ \qquad Alberto Bartesaghi$^{\mathsection}$ \qquad Guillermo Sapiro$^{\ddagger}$\thanks{Work partially supported by the ARO, ONR, NGA, AFORS, and NSF.}}

\address{
\scalebox{0.745}{
\begin{tabular}{c c c c} 
\multirow{ 2}{*}{$^{\dagger}$ Fundaci\'on Ceibal} & $^{\star}$Instituto de Ingenier\'ia El\'ectrica & $^{\mathsection}$National Cancer Institute& $^{\ddagger}$ Electrical and Computer Engineering \\
 \ & Universidad de la Rep\'ublica & National Institutes of Health & Duke University
\end{tabular}
}
}    

\begin{document}
\ninept
\maketitle
\begin{abstract}
Multi-image alignment, bringing a group of images into common register, is an ubiquitous problem and the first step of many applications in a wide variety of domains. As a result, a great amount of effort is being invested in developing efficient multi-image alignment algorithms. Little has been done, however, to answer fundamental practical questions such as: what is the comparative performance of existing methods? is there still room for improvement? under which conditions should one technique be preferred over another? does adding more images or prior image information improve the registration results? In this work, we present a thorough analysis and evaluation of the main multi-image alignment methods which, combined with theoretical limits in multi-image alignment performance, allows us to organize them under a common framework and provide practical answers to these essential questions. 
\end{abstract}

\vspace{-.35em}
\section{Introduction}
\vspace{-.4em}
Multi-image alignment consists in bringing a group of images into a common reference. It is an ubiquitous problem, being the first step of many applications, such as high dynamic range (HDR) imaging~\cite{debevec97,aguerrebere13ICCP}, super-resolution~\cite{robinson2006statistical,robinson2009optimal,liu2014bayesian}, burst deblurring~\cite{zhang2013multi,delbracio2015cvpr} and burst denoising~\cite{buades2009note}. It appears in a wide variety of domains such as computational photography, biomedical imaging, astronomy, and many other remote sensing applications, where alignment errors highly affect the final result. For instance, in computational photography applications such as HDR imaging, ghosting artifacts that appear from incorrect alignment are extremely perturbing to the observer, often voiding completely the utility of the main technique. In biomedical imaging, alignment quality often limits the resolution of structures reconstruction, detection, or segmentation~\cite{rubinstein15,grant15,bartesaghi14,li13}.

Several techniques have been proposed to tackle the multi-image alignment problem and a great amount of effort is being invested in improving results even further. 
Still, other fundamental practical aspects of the multi-image registration problem have received much less attention. For instance, what is the comparative performance of the existing methods? how much improvement can we expect if we add more images or incorporate image priors? In this work, we present a comprehensive study of the main multi-image alignment methods that allows us to organize them under a common framework and provide answers to these essential questions.
%
%
We focus on 2D rigid translations, which, despite being simple, are the basis of many models used in practice. For instance, in very complex scenes where objects move independently, we often make the sound hypothesis that the background suffers a 2D global translation, and even non-rigid transformations can be locally approximated by translations.

\mdC{Multi-image shift estimation is interlaced with the estimation of the underlying image. In fact, a straightforward and the most popular approach for recovering the latent image, is by first estimating the multiple shifts and then averaging the \emph{unshifted} noisy observations. There are other approaches that circumvent the shift estimation step and directly rely on shift-invariant features for estimating  the latent image~\cite{bendory2017bispectrum}. In this work, we focus on the problem of multiple shift estimation solely, even if as a side product we get an estimation of the underlying image. Indeed, there are many applications that estimating the shifts is an end in itself, since, for example, one could opt to combine the images later on in the image processing pipeline.}


The simplest approach to multi-image shift estimation, is to do pairwise alignment between each image in the set and one chosen as reference. However, given that all the input images share the same underlying scene, they are not independent and doing a joint alignment, mostly under low signal-to-noise ratio (SNR) conditions where the pairwise alignment is very noisy, may improve the results. The main approaches to multi-image alignment include: the maximum likelihood estimator (MLE) with different optimization strategies~\cite{robinson2009optimal,bandeira2014multireference}, the Bayesian MLE~\cite{woods06}, the maximum a posteriori (MAP) estimator with different optimization strategies~\cite{robinson2009optimal,woods06,hardie97}, and constrained alignment~\cite{govindu04,farsiu05}. 

Most of these algorithms were originally introduced for the super-resolution problem. For the multi-image alignment setting, we show  that most of these approaches are mathematically equivalent in the sense that they optimize a very similar functional, the main difference being whether they include a prior image model or not, and which is the chosen optimization technique. 

\mdC{%
In~\cite{aguerrebere16}, we presented a theoretical analysis of the fundamental limits in multi-image alignment and analyzed the performance of the MLE with respect to these bounds. Although, MLE achieves maximum performance in high SNR, there is a gap between theory and practice in medium to low SNR conditions.  In this work, we conducted a thorough experimental analysis that help us close this performance gap on a wide range of SNR conditions. 
%
%
We show that more images and an image prior are extremely useful in low SNR, enabling alignment in very challenging conditions where it is otherwise not possible. 
Indeed, the compared methods that rely on prior image information perform very close to the theoretical bounds, showing that there is little room left for improvement.


}

This article is organized as follows.  Section~\ref{sec:methods} details the analyzed methods, while Section~\ref{sec:experiment} presents an empirical evaluation on real and synthetic data. Section~\ref{sec:conclusions} summarizes the conclusions. 



\vspace{-0.8em}
\section{Multi-image Alignment Methods}
\label{sec:methods}
\vspace{-.5em}
Let us consider the image acquisition model
\begin{equation}
\z_i(\bx) = \u(\bx - \btau_i) + n_i(\bx), \quad i=0,\dots,\K, \vspace{-.5em}
\label{eq:model}
\end{equation}
where $\z_i(\bx)$ is the observed $i$-th image at pixel position $\bx = [x,y]^T$, $\u$ is the underlying continuous image, $\btau_i = [\tau_{i_x}, \tau_{i_y}]^T$ is the 2D translation vector of frame $i$ with respect to $\u$ ($\btau_0 = 0$), and $n_i(\bx)$ is independent additive Gaussian noise with variance $\sigma^2$. 
In practice, we have access to the digital images $\bz_0,\dots,\bz_K$ sampled on a discrete grid. We will assume that all the images are band-limited and sampled according to the Nyquist sampling theorem. Let $\btau = [\btau_1^T,\dots,\btau_{\K}^T]^T$ be the concatenation of all 2D unknown translations, and $\bz =\! [\bz_0^T,\dots,\bz_K^T]^T$ be the concatenation of the $(K+1)$ observed images. The goal of multi-image alignment techniques is then to estimate $\btau$ from $\bz$.  Equivalently in the Fourier domain, the model becomes: $\tilde{z}_i(\bomega) = \tilde{u}(\bomega) e^{-i \bomega \cdot \btau_i} + \tilde{n}_i(\bomega),$
%
where \texttildelow\, denotes 2D image Fourier transforms,  $\bomega = [\omega_x,\omega_y]^T$ represents the 2D Fourier spatial frequency and $\cdot$ denotes the inner product operation.


The main methods for 2D multi-image translation estimation can be classified into two categories, depending on whether they include prior image information or not. Among methods that do not include an image prior we find: the MLE, with different optimization strategies~\cite{robinson2009optimal}, and constrained alignment methods~\cite{govindu04,farsiu05}. Among those including an image prior, we find the MAP estimator, with different optimization strategies~\cite{robinson2009optimal,woods06,hardie97}, and the Bayesian MLE~\cite{woods06}. 

\vspace{.2em}

\noindent \textbf{Maximum Likelihood Estimator:} Given $(\K+1)$ independent images following Model~\eqref{eq:model}, and assuming $\u$ 
is an unknown deterministic image, the MLE of $[\u, \btau]^T$ is the value that maximizes the log-likelihood of the samples,
\begin{equation}
{[\u,\btau]}_{\textsc{mle}} = \argmax_{\u,\btau} [ \log{p(\bz | \u, \btau)} ] =  \argmin_{\u,\btau}  ||\bz - \bB(\btau) \u||^2,
\label{mleFunc}
\end{equation}
with the shift operator $\bB(\btau) = [B(\btau_1)^T,\dots,B(\btau_K)^T]^T$, where $B(\btau_i)$ is the Shannon 2D interpolation operator~\cite{woods06} that verifies: $B(\btau_i) \u(\bx) = \u(\bx - \btau_i)$. The functional in~\eqref{mleFunc} is an example of a separable non-linear least-square problem. 
Given the shifts $\btau$, the unknown underlying image $\u$ is given by the least squares solution
$\hat{\u} = (K+1)^{-1} \bB^T \bz$,
which is the average of the aligned frames. Inserting $\hat{\u}$ back into~\eqref{mleFunc}, the functional depends on the shifts only
\begin{equation}
\btau_{\textsc{mle}}  = \argmin_{\btau}  ||\bz -  (K+1)^{-1}\bB \bB^T \bz ||^2. \vspace{-.5em}
\label{mleComb}
\end{equation}
Functional~\eqref{mleComb} is non-convex and different approaches can be followed to find a local minimum~\cite{robinson2009optimal}. We consider here two approaches: cyclic coordinate descent and variable projections. Cyclic coordinate descent consists in optimizing~\eqref{mleComb} on one coordinate at a time. Two main steps are iterated: first, compute the average of the frames aligned with the current estimate of the shifts (given by $\hat{\u}$); second, find each shift by minimizing the Euclidean distance of the corresponding image against the average.
Variable projections makes use of the fact that minimizing~\eqref{mleComb} is equivalent to 
\begin{equation}
\btau_{\textsc{mle}_{vp}}  = \argmax_{\btau} \bz^T \bB\ \bB^T \bz. \vspace{-.5em}
\label{eq:vpMLE}
\end{equation}
Robinson \etal{}~\cite{robinson2009optimal} proposed to compute the MLE maximizing~\eqref{eq:vpMLE}, a method they named variable projections. For the super-resolution problem, they claim that this optimization has several advantages compared to the cyclic coordinate descent approach since it converges in fewer iterations and its minima are better defined. 




\vspace{.2em}

\noindent \textbf{Bayesian Maximum Likelihood Estimator:} 
Let us assume that the underlying image $\u$ can be modeled as a stationary zero-mean Gaussian process with spectral density $\bS_{\u}=\bF^T\bSigma_{\u}\bF$ , where  $\bSigma_{\u}$ is the covariance matrix and $\bF$ the Fourier operator. Under this hypothesis, $\u$ can be considered a hidden variable and be marginalized from the samples log-likelihood. 
The unknown shifts can then be computed maximizing the marginal likelihood $p(\bz | \btau) = \int p(\bz| \u, \btau) p(\u) \mathrm{d} \u$, which is also a Gaussian function (see e.g., \cite[Eq. (2.115)] {bishop2006}), having zero mean and covariance matrix $\bGamma = \sigma^2 \bI + \bB\bSigma_u\bB^T$. Then, 
\begin{equation}
\begin{aligned}
\btau_{\textsc{bmle}}	 &= \argmax_{\btau} \tilde{\bz}^T \tilde{\bB} \tilde{\bW} \tilde{\bB}^T \tilde{\bz},
\label{eq:vpMLEB}
\end{aligned}
\end{equation}
where $\tilde{\bB} = \bF \bB \bF^{T}$ is the Fourier equivalent of the shift operator, $\tilde{\bW} = ( \bS_{\u} (K+1) + \bI \sigma^2)^{-1} \bS_{\u}$ is the Wiener filter, and $\bI$ is the identity matrix. 
Notice that the only difference between the Bayesian maximum likelihood estimator (BMLE)~\eqref{eq:vpMLEB} and the MLE~\eqref{eq:vpMLE} is the Wiener filter, which appears with the introduction of the image prior. 



\vspace{.2em}

\noindent \textbf{Maximum a Posteriori:}
An alternative way to incorporate an image prior is to compute the MAP estimator. Given $(\K+1)$ independent samples following Model~\eqref{eq:model}, and assuming $\u$ 
is drawn from a zero-mean Gaussian process with spectral density $\bS_{\u}$, the MAP of $[\u, \btau]^T$ is the value that maximizes the posterior probability
\begin{equation}
\begin{aligned}
{[\u,\btau]}_{\textsc{map}} &= \argmax_{\u,\btau} [ \log{p(\bz | \u, \btau)} + \log{p(\u)}] \\
     &= \argmin_{\u,\btau} \tfrac{1}{2\sigma^2}||\bz - \bB^T \u ||^2 + \tfrac{1}{2}\u^T \bSigma_{\u}^{-1} \u.
\label{eq:MAP1}
\end{aligned}
\end{equation}
Similarly to functional~\eqref{mleFunc}, functional~\eqref{eq:MAP1} is an example of a separable non-linear least-square problem. Following the same steps as for the MLE, it can be shown that minimizing~\eqref{eq:MAP1} is equivalent to~\cite{robinson2009optimal}
\begin{equation}
\btau_{\textsc{map}}  = \argmax_{\btau} \tilde{\bz}^T \tilde{\bB} \tilde{\bW} \tilde{\bB}^T \tilde{\bz},\vspace{-.5em}
\label{eq:MAP}
\end{equation}
where $\tilde{\bW}$ is the Wiener filter as defined before. Hence, the MAP and the BMLE optimize the same cost function. 

\vspace{.3em}

\noindent \textbf{Common Framework.}
While often presented as different techniques, we have just showed that the previously presented multi-image alignment approaches, all optimize the cost function\footnote{This equivalence is not necessarily valid when addressing the super-resolution problem (including blurring and subsampling operatiors).}
$
E(\btau) = \tilde{\bz}^T \tilde{\bB} \tilde{\bW} \tilde{\bB}^T \tilde{\bz},
$
whether with $\tilde{\bW} = \bI$ for the methods without image prior or $\tilde{\bW}$ equal to the Wiener filter for methods including an image prior. Nevertheless, given that this functional is non convex, results may vary depending on the optimization strategy  and the initialization procedure. 

\vspace{.3em}

\noindent \textbf{Constrained Alignment.} Another way of aligning multiple images is to use all possible pairwise estimations and use their redundancy to get more accurate and self-consistent shifts~\cite{govindu04}. For this purpose, all the pairwise shifts $b_{ij}$ between any two frames $i < j$ are first computed by locating the maximum value of their correlation map. Then, these estimates are combined making use of the fact that $b_{ij}$ is the summation of the shift vectors of all intermediate adjacent frames,
$
r_i + r_{i+1} + \dots + r_{j-1} = b_{ij}
$.
The shifts $r_i$ can then be found solving the over-determined set of linear equations ($K(K+1)/2$ equations and $K$ unknowns) obtained by determining all possible shifts between any two frames. A similar approach was proposed by Farsiu \etal{}~\cite{farsiu05}, where they incorporate the coherence as a constraint in the optimization of the shifts directly.



\vspace{.3em}
\noindent \textbf{Theoretical Performance Limits.}
In~\cite{aguerrebere16} we presented a theoretical study deriving different statistical performance bounds for the translations estimation accuracy in multi-image alignment. The Cram\'er-Rao bound, under the assumption that the underlying image follows a natural image prior (CRB) was derived. The CRB gives a lower bound on the mean square error (MSE) of any unbiased estimator of the shifts $\btau$ (see \cite[Eq. (44)]{aguerrebere16}). 
%
%
Different behaviors for the alignment accuracy are identified, depending on the SNR of the input images. For very high SNR, the performance bound is independent of the number of images $K$ or prior information (image or shifts prior), and the MLE attains the bound. Doing pairwise alignment using the MLE is optimal in this case. For high to moderate SNR, increasing the number of images does improve performance. Interestingly, theory predicts the existence of an SNR threshold below which performance degrades briskly and a lower limit SNR value below which alignment is not possible. Nevertheless, increasing the number of images can push back these thresholds several dBs, making alignment possible in much more challenging conditions. 
For natural images, the MLE performance is close to the CRB but it is not tight. A possible reason for this is the non-optimality of the MLE, for which a critical drawback is that it does not use image prior information. Moreover, for this SNR region, the MLE performance clearly improves with increasing
number of images.  Including prior image information, for example through a Bayesian approach such as minimizing the expected MSE or computing the Maximum a Posteriori, could help close the gap between the fundamental limits and the MLE performance.
%
%
%

\begin{figure}
\centering
\begin{minipage}[c]{0.18\textwidth}
\centering
\includegraphics[clip=true,trim={0 10 0 5},width=0.4\linewidth]{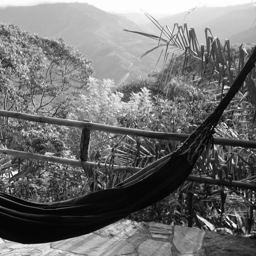}
\includegraphics[clip=true,trim={0 10 0 5},width=0.4\linewidth]{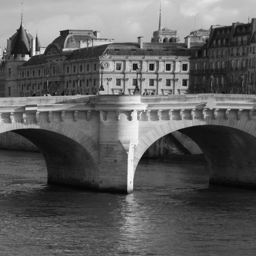}\\
\includegraphics[clip=true,trim={0 10 0 5},width=0.4\linewidth]{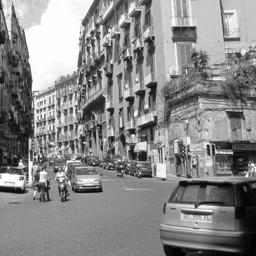}
\includegraphics[clip=true,trim={0 10 0 5},width=0.4\linewidth]{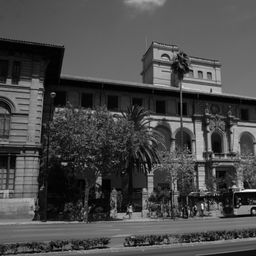}\\
\vspace{1pt}
\scriptsize{(a)}
\end{minipage}
\begin{minipage}[c]{0.22\textwidth}
\centering
\includegraphics[width=\linewidth,height=7.75em]{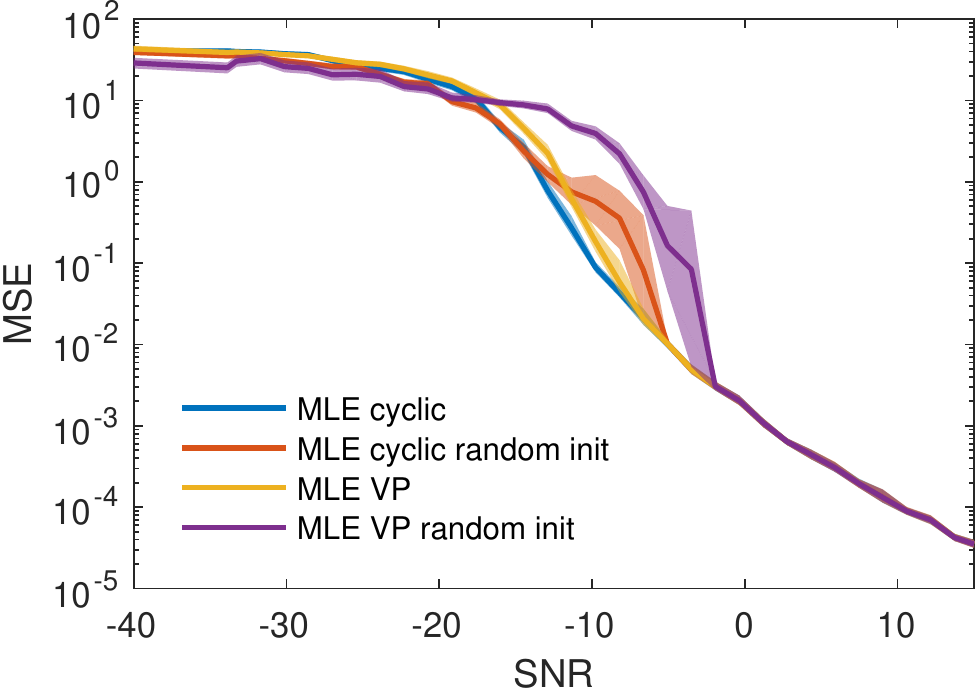}
\vspace{-2pt}
\scriptsize{(b)}
\end{minipage}
\begin{minipage}[c]{0.22\textwidth}
\centering
\includegraphics[width=\linewidth,height=7.75em]{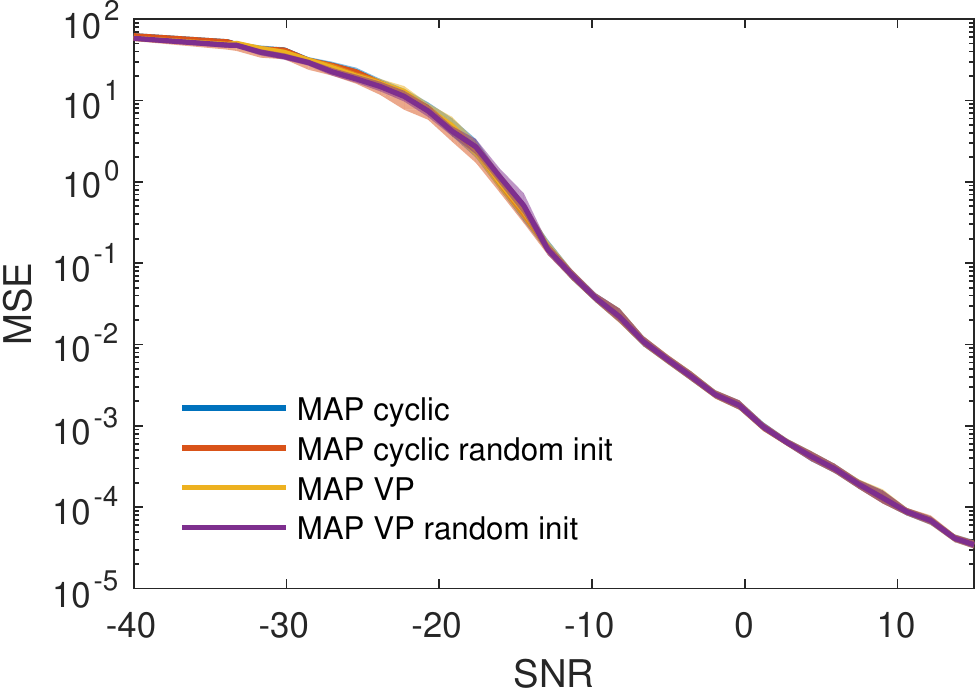} 
\vspace{-2pt}
\scriptsize{(c)}
\end{minipage}
\begin{minipage}[c]{0.22\textwidth}
\centering
\includegraphics[width=\linewidth,height=7.75em]{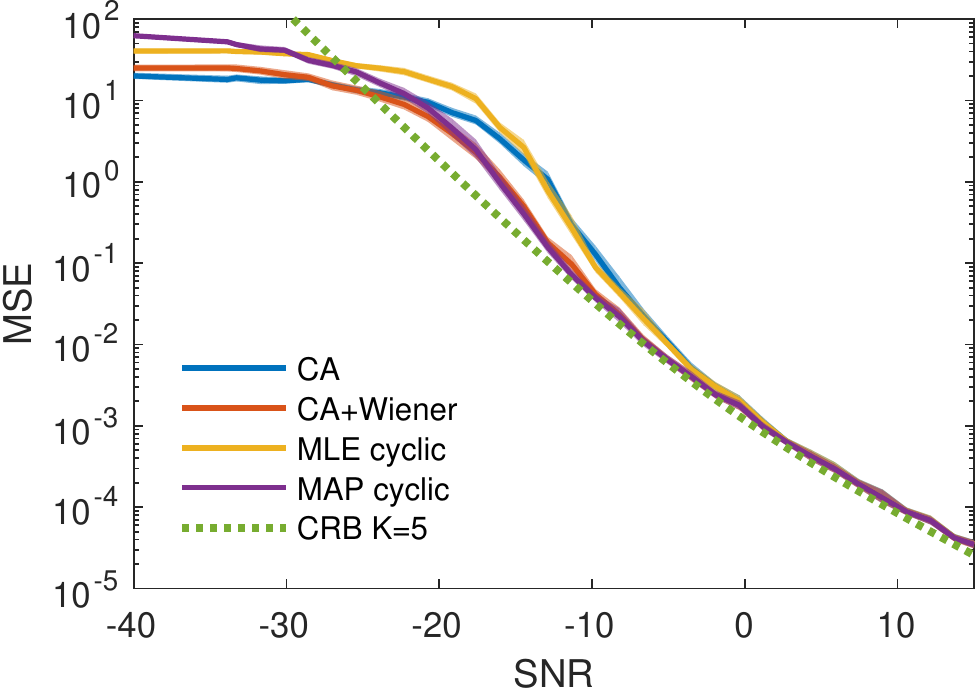} 
\vspace{-2pt}
\scriptsize{(d)}
\end{minipage}
\vspace{-.5em}
 \caption{(\textbf{a}) Natural images used as ground-truth (images from~\cite{aguerrebere16}). Comparison of optimization and initialization strategies of MLE (\textbf{b}) and MAP (\textbf{c}) with $K=5$ images. (\textbf{d}) Performance comparison of MLE, MAP and the constrained alignment applied to Wiener filtered images. The CRB is displayed as a performance benchmark~\cite{aguerrebere16}.}
\label{fig:optInit}
\end{figure}

\vspace{-.5em}
\section{Experimental Analysis}
\label{sec:experiment}
\mdC{In what follows, we conduct a thorough experimental analysis  to evaluate the presented techniques with both synthetic and real data.}



\vspace{-.8em}
\subsection{Synthetic Data}
\label{sec:exps}

\vspace{-.3em}
\noindent \textbf{Experimental setup:} Sets of images are generated following Model~\eqref{eq:model}, taking as ground-truth the examples in Fig.~\ref{fig:optInit} (a) ($50 \times 50$ pixels), with different noise levels and number of shifted images. Two types of motion are considered: uniformly distributed independent shifts and drift-driven trajectories (each image is shifted from the previous image position, according to prior angle and speed distributions), often observed in biomedical applications (e.g., cryo-electron microscopy) and image bursts capture with hand-held cameras, among others. The image prior for the MAP estimator is a stationary zero-mean Gaussian process with a spectral density that decays as the inverse of the squared Fourier frequency, a prior widely used for locally modeling natural images~\cite{aguerrebere16,elad2001fast,fransens2007optical,levin2009understanding,efrat2013accurate}. The experiments are repeated 100 times for each SNR level and the mean and  $95\%$ confidence intervals of the MSE are reported. The methods are almost unbiased (the squared bias was on average, in all experiments, three orders of magnitude smaller than the variance). The SNR is defined as the ratio between the total energy of the derivative of the ground-truth image and the noise power~\cite{aguerrebere16}. 

\vspace{.2em}

\noindent \textbf{Optimization and initialization:} The MLE and MAP were solved using both cyclic coordinate descent and variable projections. All optimizations were initialized using random shifts as well as pairwise alignment (each image is aligned to the first image using classical correlation). Fig.~\ref{fig:optInit} shows the results for the MLE (b) and MAP (c) with $K=5$ images with a random shifts trajectory. For the MLE, in high SNR conditions, there are no significant performance differences between the optimizations nor the initialization strategies. For lower SNR levels, however, the cyclic coordinate descent optimization seems to be slightly better than variable projections, and the pairwise alignment initialization considerably outperforms the random initialization. For the MAP approach, no significant differences are found between the tested configurations at all SNR levels. 


\begin{figure}

\begin{minipage}[c]{0.45\linewidth}
\centering
\includegraphics[width=\linewidth,height=7.75em]{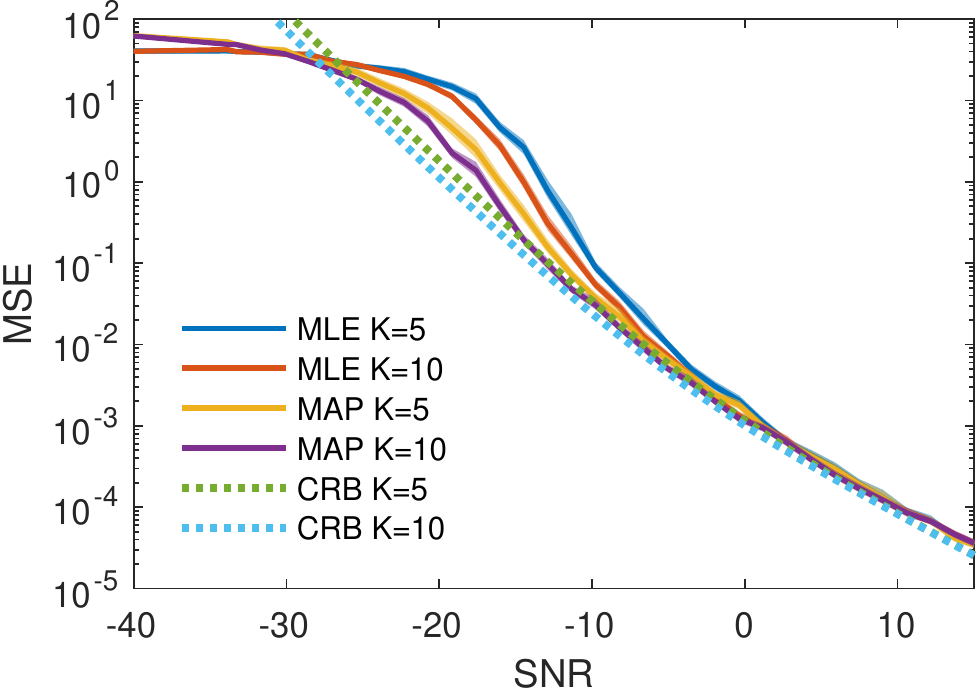}
\scriptsize{(a)}
\end{minipage}
\begin{minipage}[c]{0.45\linewidth}
\centering
\includegraphics[width=\linewidth,height=7.75em]{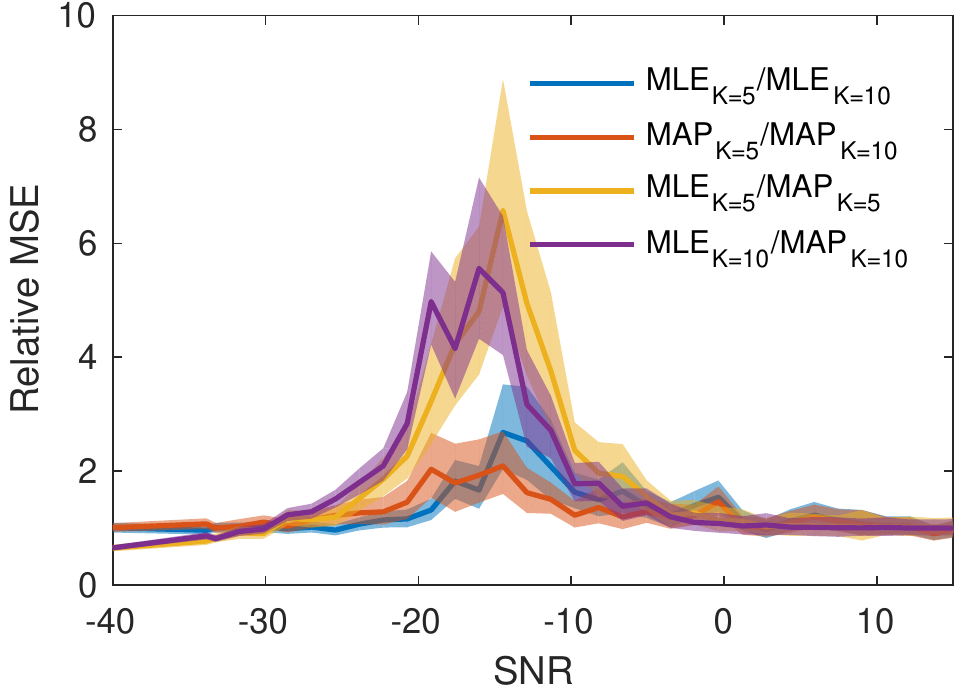}  
\scriptsize{(b)}
\end{minipage}
\vspace{-.4em}
\caption{ (\textbf{a}) MLE versus MAP performance as a function of the input image SNR when using $K=5$ and $K=10$ images, and the respective CRB bounds. (\textbf{b}) Ratio of the performance of the different configurations shown in (a). }
\label{fig:imagePrior}
\end{figure}

\vspace{.2em}

\noindent \textbf{Image prior:} Fig.~\ref{fig:optInit} (d) compares the performance of the cyclic coordinate descent MLE and MAP, both initialized using pairwise alignment, and the constrained alignment estimator applied to Wiener prefiltered images (assuming the same image prior as for MAP). The results are compared to the Cram\'er-Rao lower bound (CRB)~\cite[Eq. (44)]{aguerrebere16}, which gives a lower bound on the variance of any unbiased estimator of the shifts, therefore establishing a performance benchmark. The CRB is computed assuming the same image prior as for the MAP estimator (see~\cite{aguerrebere16} for details). 
%
For very high SNR, all methods perform very similarly and close to the CRB, meaning that the extra information provided by the image prior is not useful. For low to moderate SNR levels, however, a clear performance improvement is observed with the MAP estimator which reduces the gap between the MLE and the limit predicted by the CRB. Even more importantly, including the image prior pushes back several dBs the SNR threshold after which alignment performance degrades dramatically (about -10dB for MLE to -14dB for MAP, see Fig.~\ref{fig:optInit} (d)). Hence, including the image prior enables alignment in very challenging noise conditions where it is otherwise not possible. 
\vspace{.3em}

\noindent \textbf{Number of images:} Fig.~\ref{fig:imagePrior} shows a comparison of MLE and MAP with different number of images ($K=5,10$). Similarly to what was observed for the image prior, increasing the number of images has no effect in high SNR conditions. For lower SNR, increasing the number of images improves the results, both for MLE and MAP. It is interesting to remark, however, that the performance gain given by including the image prior is much larger than that of increasing the number of images. Indeed, the MAP estimator with 5 images performs considerably better than the MLE with 10 images. Fig.~\ref{fig:imagePrior} (b) shows the ratio between the performance of the different configurations. Increasing the number of images produces a higher improvement for MLE (blue curve) than for MAP (red curve).  But more importantly, the improvement is much larger when including the image prior (MLE/MAP ratio shown in the yellow and violet curves), showing a larger gain for smaller $K$. This is particularly interesting from a practical perspective given the increased complexity, and thus time requirements, of increasing the number of images as opposed to the almost costless inclusion of the appropriate image prior.


The previous results correspond to \emph{bolivia} image (Fig.~\ref{fig:optInit} (a) top left) with independent random shifts, but the same behavior is observed for the other examples and for drift-driven trajectories as well. 
\begin{figure}
\centering
\begin{minipage}[c]{.45\linewidth}
\centering
\scriptsize{no image prior (MLE)}\\
\includegraphics[trim={0 60 0 40},clip,width=\linewidth]{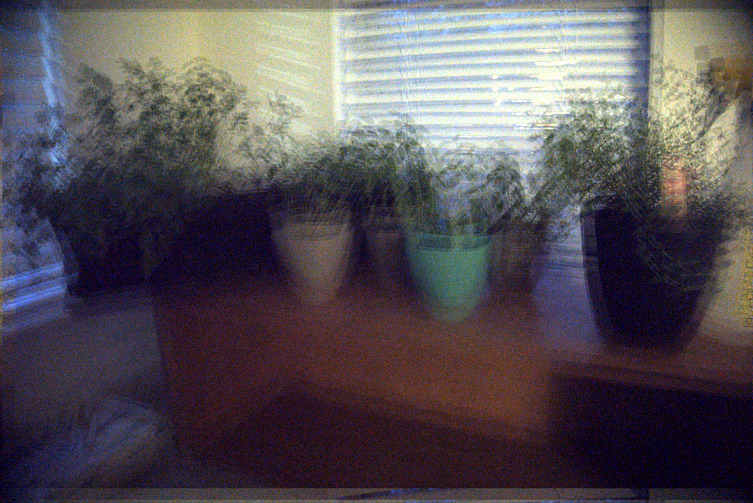}
\end{minipage}
\begin{minipage}[c]{.45\linewidth}
\centering
\scriptsize{with image prior (MAP)}\\
\includegraphics[trim={0 60 0 40},clip,width=\linewidth]{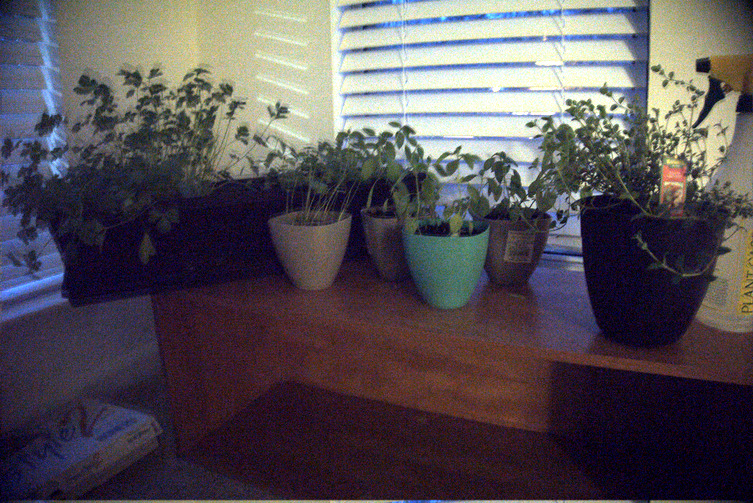}
\end{minipage}


\vspace{.3em}

\begin{minipage}[c]{.01\linewidth}
\rotatebox{90}{\hspace{15pt} \scriptsize{no prior}} \rotatebox{90}{\hspace{10pt} \scriptsize{prior}}
\end{minipage}
\hspace{.125em}
\begin{minipage}[c]{.91\linewidth}
\centering
\begin{minipage}[c]{0.19\textwidth}
\includegraphics[trim={0 10 0 20},clip,width=\textwidth]{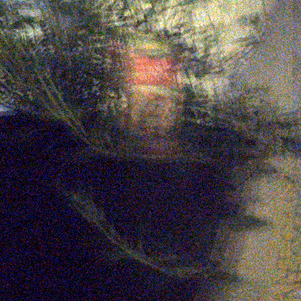}
\end{minipage}
\hspace{-4pt}
\begin{minipage}[c]{0.19\textwidth}
 \includegraphics[trim={0 10 0 20},clip,width=\textwidth]{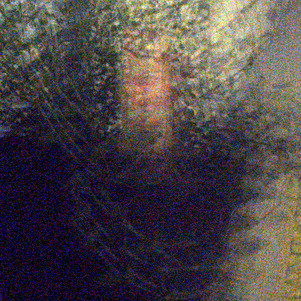} 
\end{minipage}
\hspace{-4pt}
\begin{minipage}[c]{0.19\textwidth}
\includegraphics[trim={0 10 0 20},clip,width=\textwidth]{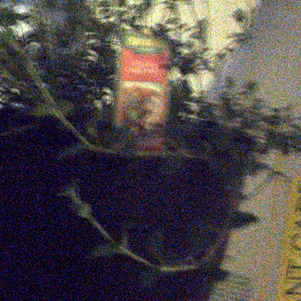} 
\end{minipage}
\hspace{-4pt}
\begin{minipage}[c]{0.19\textwidth}
\includegraphics[trim={0 10 0 20},clip,width=\textwidth]{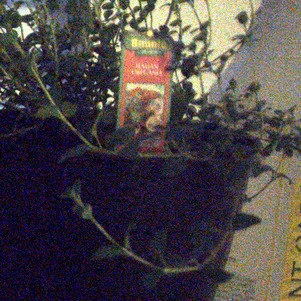} 
\end{minipage}
\hspace{-4pt}
\begin{minipage}[c]{0.19\textwidth}
\includegraphics[trim={0 10 0 20},clip,width=\textwidth]{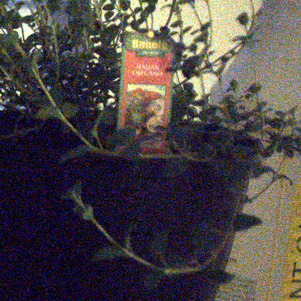} 
\end{minipage}


\begin{minipage}[c]{0.19\textwidth}
\includegraphics[trim={0 10 0 20},clip,width=\textwidth]{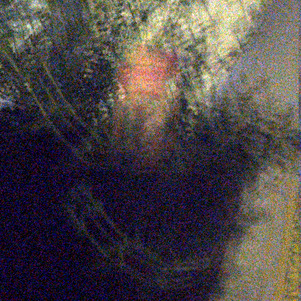}
\end{minipage}
\hspace{-4pt}
\begin{minipage}[c]{0.19\textwidth}
\includegraphics[trim={0 10 0 20},clip,width=\textwidth]{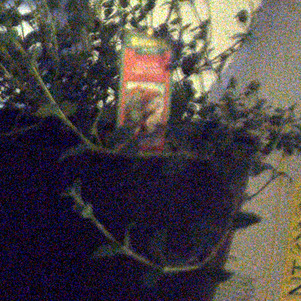} 
\end{minipage}
\hspace{-4pt}
\begin{minipage}[c]{0.19\textwidth}
\includegraphics[trim={0 10 0 20},clip,width=\textwidth]{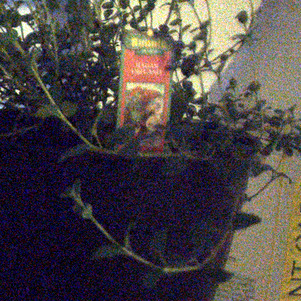} 
\end{minipage}
\hspace{-4pt}
\begin{minipage}[c]{0.19\textwidth}
\includegraphics[trim={0 10 0 20},clip,width=\textwidth]{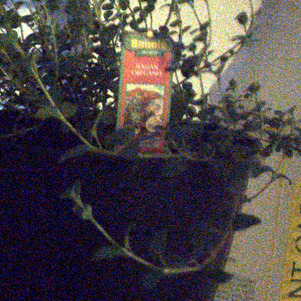} 
\end{minipage}
\hspace{-4pt}
\begin{minipage}[c]{0.19\textwidth}
\includegraphics[trim={0 10 0 20},clip,width=\textwidth]{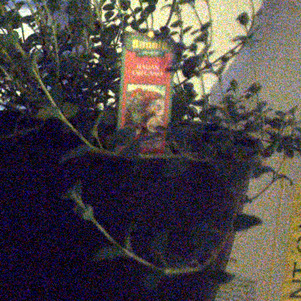} 
\end{minipage}

%


\begin{minipage}[c]{0.19\textwidth}
\centering
\scalebox{0.8}{\scriptsize{$64\times 64$}}
\end{minipage}
\hspace{-4pt}
\begin{minipage}[c]{0.19\textwidth}
\centering
\scalebox{0.8}{\scriptsize{$128\times 128$}}
\end{minipage}
\hspace{-4pt}
\begin{minipage}[c]{0.19\textwidth}
\centering
\scalebox{0.8}{\scriptsize{$256\times 256$}}
\end{minipage}
\hspace{-4pt}
\begin{minipage}[c]{0.19\textwidth}
\centering
\scalebox{0.8}{\scriptsize{$512\times 512$}}
\end{minipage}
\hspace{-4pt}
\begin{minipage}[c]{0.19\textwidth}
\centering
\scalebox{0.8}{\scriptsize{$1024 \times 1024$}}
\end{minipage}
\end{minipage}
\vspace{-.4em}
\caption{\textbf{Top:} Comparison of the alignment with and without prior when doing the shift estimation using a patch size $128\times 128$ and $K=5$ images. \textbf{Bottom:} Alignment with increasing patch size (left to right) without (top) and with (bottom) image prior.} 
\label{fig:compareReal}
\end{figure}

\vspace{-1.1em}
\subsection{Real Data}
\label{ssec:realDataExp}
To evaluate the applicability of the previous results to real data, we compare the alignment performance for the burst denoising task~\cite{buades2009note}. For burst denoising (a widely used and very powerful technique for noise reduction in low light conditions) correct alignment is essential to obtain good ghosting-free results. 

\vspace{.3em}

\noindent \textbf{Experimental setup:} Image bursts are acquired in low light conditions with a hand-held Sony $\alpha$-5100 camera set to ISO 20000. Burst denoising is performed by aligning and averaging groups of $K=2,5,10,$ raw images. Images are aligned using the MLE and MAP solved by cyclic coordinate descent, considering the same image prior as before. 
To have different SNR levels, alignment is performed using a sub-image (patch) of varying size: $64 \times 64,\ldots, 1024 \times 1024$. The gradient content increases with the patch size, thus reproducing conditions of increasing SNR. 
The shifts are estimated using the red channel, then applied to all channels before demosaicking.

\vspace{.3em}

\begin{figure}
\begin{minipage}[c]{0.6\linewidth}
\centering
\includegraphics[trim={0 30 0 30},clip,width=0.3\textwidth]{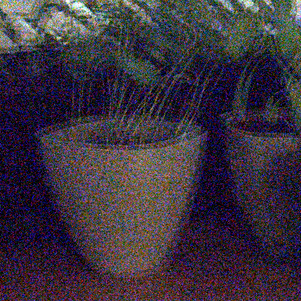}\hspace{0.5pt}
\includegraphics[trim={0 30 0 30},clip,width=0.3\textwidth]{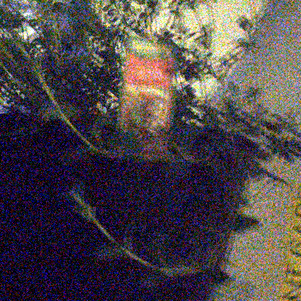}\hspace{0.5pt}
\includegraphics[trim={0 30 0 30},clip,width=0.3\textwidth]{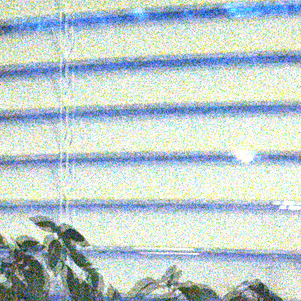}

\includegraphics[trim={0 30 0 30},clip,width=0.3\textwidth]{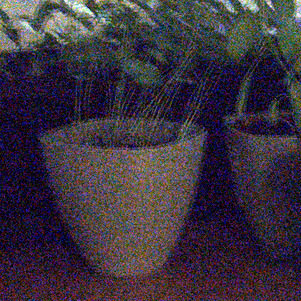}\hspace{0.5pt}
\includegraphics[trim={0 30 0 30},clip,width=0.3\textwidth]{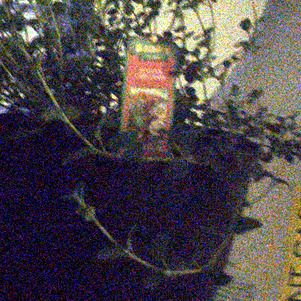}\hspace{0.5pt}
\includegraphics[trim={0 30 0 30},clip,width=0.3\textwidth]{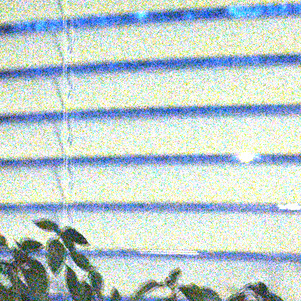}

\end{minipage}
\begin{minipage}[c]{0.38\linewidth}
\centering
\includegraphics[width=\textwidth]{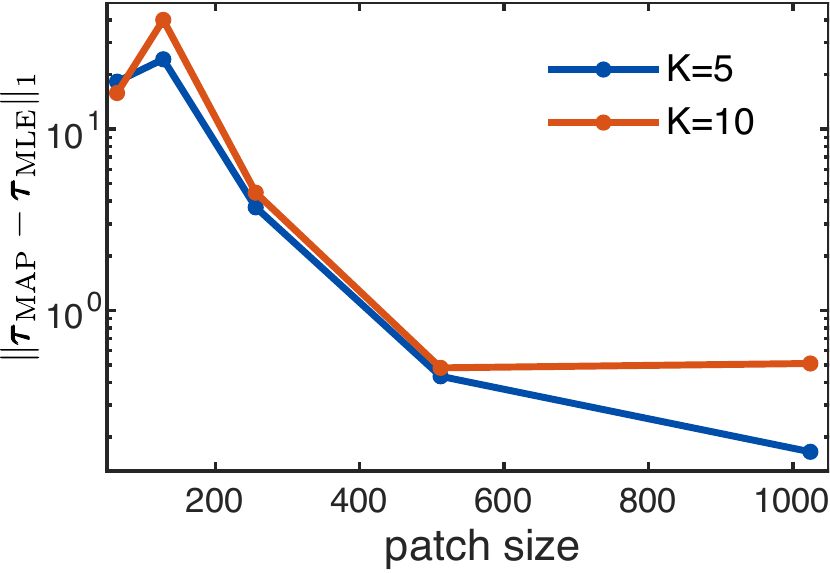}
\end{minipage}
\vspace{-.6em}
\caption{\textbf{Left:} Extracts of the average of two images aligned using the shifts obtained with $K=2$ (pairwise) and $K=10$ images (top and down respectively). \textbf{Right:} Difference between MAP and MLE estimations for varying patch size and number of images.}
\label{fig:compareReal1}
\end{figure}

\noindent \textbf{Effects of the image prior:} Fig.~\ref{fig:compareReal} (top) shows an example of the results obtained with 5 images, where alignment is performed using a sub-region of $128 \times 128$ pixels. In this case, alignment is not possible without the image prior but it becomes feasible including it.
Fig.~\ref{fig:compareReal} (bottom) shows an extract of the results obtained when using different patch sizes. The first and second rows show the result without and with image prior, respectively. 
For the two smallest patch sizes ($64 \times 64$ and $128 \times 128$) the noise is very high and there is almost no gradient in the patch, representing cases of very low SNR. Even under these very challenging conditions, the image prior enables alignment with the $128 \times 128$ patch. This result confirms what was observed in Section~\ref{sec:exps}, which showed that the SNR threshold below which alignment performance degrades dramatically can be pushed back by including an image prior. For the patch size $256 \times 256$, the gradient content is already visible and alignment is possible with or without the image prior. Nevertheless, the quality difference is still clear. For patch sizes above $512$, the difference is almost indistinguishable. This was also predicted by the results obtained in Section~\ref{sec:exps}, since in high SNR all methods agree and attain the CRB~\cite{aguerrebere16}. 
Fig.~\ref{fig:compareReal1} (right) shows the absolute value of the difference between the shift estimates obtained with and without image prior for the different patch sizes and different number of images (average of the difference in $x$ and $y$ directions). The estimations are very similar for patch sizes above 512. 
%

\noindent \textbf{Effects of the number of images:} Fig.~\ref{fig:compareReal1} (left) compares the quality of the estimated shifts with MAP using 2 (pairwise alignment) versus 10 images for a patch size of $128 \times 128$. For this purpose we show the average of two images aligned using the shifts obtained with $K=2$ (left) and using $K=10$ (right). The result of the multi-image alignment is sharper than the pairwise alignment, showing the clear performance improvement obtained by using more images.

\vspace{-.5em}
\section{Conclusions}
\vspace{-.5em}
\label{sec:conclusions}
We conducted a thorough analysis on multi-image shift estimation methods. 
We showed that most of them use different optimization techniques to optimize the same functional, 
the main difference being the inclusion of prior image information. 
We then conducted an experimental analysis, that confirmed the per-region behavior depending on the SNR conditions predicted by a theoretical analysis on the fundamental limits of multi-image alignment performance~\cite{aguerrebere16}. 
In very high SNR, all the evaluated methods perform very similarly and very close to the Cram\'er-Rao lower bound. 
Hence, the simplest methods already achieve the best possible performance in this SNR condition and including prior image information or more images does not improve the alignment result. 
For moderate to low SNR, however, we show that there is a clear performance gain when including an image prior or using more images. This gain is twofold: the MSE is reduced and, more importantly, the threshold at which performance degrades dramatically is pushed back several dBs. Therefore, for these SNR conditions, including more images or an image prior makes alignment possible in conditions where it is otherwise not possible. The performance gain obtained by including the image prior is larger than that of increasing the number of images. 
This is important from a practical perspective given the increased complexity, and therefore time requirements, of increasing the number of images as opposed to the almost costless inclusion of the image prior. Indeed, we found that the methods that include the image prior perform very close to the CRB for a larger SNR range, showing that there is little room left for improvement as they close the gap between the MLE and the CRB~\cite{aguerrebere16} in moderate to low SNR. 
Regarding optimization and initialization, slight differences are observed for the MLE in low SNR conditions, but this does not seem to be a critical aspect. Finally, as predicted by theory~\cite{aguerrebere16}, we observe the existence of an SNR threshold below which none of the evaluated methods manages to align the images, and neither more images nor an image prior can revert this situation. The only way out is increasing the SNR, e.g., increasing the image size (or patch size in case of local alignment).

\bibliographystyle{IEEEbib}

\bibliography{references}
\end{document}